# HIGH PERFORMANCE PARALLEL SORT FOR SHARED AND DISTRIBUTED MEMORY MIMD

Thoria Alghamdi[1] and Gita Alaghband[1]

*Department of Computer Science and Engineering, University of Colorado Denver -Denver, CO, 80217 [1]*

**ABSTRACT**

We present four high performance hybrid sorting methods developed for various parallel platforms: shared memory multiprocessors, distributed multiprocessors, and clusters taking advantage of existence of both shared and distributed memory. Merge sort, known for its stability, is used to design several of our algorithms. We improve its parallel performance by combining it with Quicksort. We present two models designed for shared memory MIMD (OpenMP): (a) a non-recursive Merge sort and (b) a hybrid Quicksort and Merge sort. The third model presented is designed for distributed memory MIMD (MPI) using a hybrid Quicksort and Merge sort. Our fourth model is designed to take advantage of the shared memory within individual nodes of today's cluster systems, and to eliminate all internal data transfers between different nodes, Our model implements a one-step MSD-Radix to distribute data in ten packets (MPI) while parallel cores of each node use Quicksort to sort their data partitions sequentially then merge and sort them in parallel employing the OpenMp. The performances of all developed models outperform the baseline performance. Hybrid Quicksort and Merge sort outperformed Hybrid Memory Parallel Merge Sort using Hybrid MSD-Radix and Quicksort in Cluster Platforms when sorting small size data, but with larger data the speedup of Hybrid Memory Parallel Merge Sort Using Hybrid MSD-Radix and Quicksort in Cluster Platforms becomes bigger and it keeps improving. The speedup of Distributed Memory Parallel Hybrid Quicksort and Merge Sort is the best.

**KEYWORDS**

Merge Sort, MPI, OpenMP, Parallel Sort, Quick Sort, Radix Sort.

## 1. INTRODUCTION

Sorting data is an essential process in computation. Numerous applications are built on or rely upon sorting: searching, shuffling, databases, transaction processing, scientific applications, string matching to just name a few (*Aydin & Alaghband 2013*, *Zurek* et al. *2013).* This has led to many sequential fast and efficient sorting algorithms. Parallel sorting algorithms pose two main challenges. First, implementing sorting algorithms in a parallel platform is not straightforward (*Durad 2014*, *Zurek* et al. *2013).* In a parallel environment, bookkeeping code to determine indices of elements to be compared and swap by parallel processes incur a large computation compared with the simple compare and swap of the elements being sorted. If the bookkeeping is done sequentially, or the size of data to be sorted is small the performance will degrade below that of a sequential sort. Second, the overhead associated with communication needed between processes can be larger than the performance gain from penalization. The parallel sorting algorithm must therefore be designed with these constraints in mind.

The first goal of this paper is to implement fast parallel hybrid Merge sort and Quick sort in shared memory and in distributed memory using OpenMP and MPI. The second goal is to implement the parallel hybrid Merge sort and Quick using one step of MSD-Radix before dividing data between nodes in cluster platform.

Section 2 gives a brief introduction to Merge sort, Quicksort, and Radix sort algorithms. Also, it shows a general overview of Shared and Distributed Memory and it includes related works. Implementation details of Shared Memory and Distributed Memory of Parallel Hybrid Merge sort and Quicksort algorithms, and Hybrid Memory Parallel Merge Sort Using Hybrid MSD-Radix and Quicksort in Cluster Platforms are explained in Section 3. Experiments and performance analysis of proposed algorithms are given in Section 4.

# BACKGROUND

## 2.1 Sorting Algorithms Overview

This paper focuses on three fast sorting algorithms; Merge Sort, Quicksort, and Radix Sort. Each algorithm shows strength in a specific area. Merge sort and Quicksort are considered as divide-and-conquer methods (*Al-Dabbagh & Barnuti 2016, Radenski 2011*). Their average complexity is *O(nlogn)* but practical implementation showed slower performance for Merge sort than Quick sort (*Al-Dabbagh & Barnuti 2016*). Merge sort is a stable sort algorithm, but it requires additional memory equal the size of the input (*Al-Dabbagh & Barnuti 2016*). So, to improve the parallel Merge sort algorithm, we implemented Quicksort algorithm sequentially since experiment showed it is faster than sequential Merge sort. Then we implemented merge algorithm in parallel between threads or nodes. Radix sort algorithm can be implemented as either least-significant-digit (LSD) or the most-significant-digit (MSD). For data of integer type, Radix uses ten buckets to store data depending on their digit values (*Aydin & Alaghband 2013*). Radix is a proper algorithm to divide data according to their most significant digit which makes it the best choice to divide data between nodes and guarantee that there is no overlap between nodes. By using MSD-Radix, we can decrease the communication between nodes.

## 2.2 Shared Memory and Distributed Memory MIMD

In shared memory MIMD, processors read and write using their normal memory reference instructions (*Jordan & Alaghband 2003)*. These systems are considered easier for application design and programming perspective than the distributed MIMD computers because all processors can access data directly and share the same logical address space. Using single address space reduces the problems of data partitioning, migration and load balancing. OpenMP is a parallel programing language designed for shared memory MIMD (*Jordan & Alaghband 2003)*. The shared memory also improves the ability of parallelizing compilers, standard operating systems, resource management and incremental performance tuning of applications (*Kavi et al 2000*). However, there is a cost associated with parallelism using OpenMp. This cost represents the overhead of threads management including the time to create, start, and stop threads, the time to assign the work to the threads, and the time spent for synchronization *(Chapman et al 2007).*
In distributed memory MIMD, processors access local memory directly, and use messages system to access remote data in other processors (*Kavi et al 2000, Jordan & Alaghband 2003*). MPI is a parallel programing language designed for distributed memory MIMD which can also run on shared memory MIMD computers *(El-Nashar 2011, Jordan & Alaghband 2003)*. A cluster consists of several distributed compute nodes, each configured as a shared memory MIMD. The memory in each node is shared among processors (cores) of the node, but is local for all processors on other nodes. Programming paradigm supported by such a cluster, can be either completely distributed (MPI) or a hybrid of distributed and shared memory (MPI/OpenMP).

## 2.3 Related Works

To gain performance from parallel sort, researchers used different strategies; one strategy is to improve the sequential sort algorithm then employ it in the parallel algorithm such as turning recursive sort algorithm to non-recursive (*Aydin & Alaghband 2013). A*nother strategy is to mix two or more sorting algorithms in one parallel algorithm. This strategy becomes possible in parallel platform. Study shows that using Quicksort with MSD-Radix in a hybrid parallel algorithm gives better performance than using MSD-Radix alone (*Aydin & Alaghband 2013).* Other strategy is to use hybrid memory in cluster platform to get the advantage of shared memory inside the node in distributed memory. Other researchers investigated the effect of data structure on the performance of the parallel sort algorithm. Study shows that using array in sorting code gives better performance than using vector (*Alyasseri et al. 2014*).

# SORTING ALGORITHMS IMPLEMENTATION

In this study, we implemented three sequential sort algorithms and four parallel sort algorithms. We select from the best performing efficient sequential sorts (Quicksort, recursive and non-recursive Merge sort) to design and compare two parallel sorting algorithms for shared memory MIMD (non-recursive Merge and Hybrid Quicksort and Merge Sort) using OpenMP. The best performing hybrid model from this result is used to design and implement an efficient distributed sort (Hybrid Quicksort and Merge Sort) using MPI. The fourth implementation is designed for a cluster with hybrid memory that uses MSD-Radix sort between the nodes (MPI) which reduces the communication overhead in the system and Quicksort within each compute node (OpenMP) with the best performance.

## 3.1 Sequential Merge Sort and Quicksort

To implement sequential recursive Merge sort, we use two functions. First function is for splitting the data recursively till size two elements or less. The second function merges the two sorted lists in one sorted list *(Hijazi & Qatawneh 2017)*. The non-recursive Merge sort. sorts each two elements together, then uses the same concept of merging to merge two sorted lists and repeats for each two lists until all elements are sorted. To implement the Quicksort algorithm, we choose an element as a pivot, then sort around the pivot to set all smaller numbers to the right and all greater numbers to the left of the pivot. Quicksort is called for each list repeatedly until all elements are sorted *(Kim & Park 2009)*. Figure 1 shows sorting algorithms pseudocode to sort array *S* of size *n* (a) shows the recursive Merge sort algorithm, (b) shows the non-recursive Merge sort algorithm, and (c) shows the recursive Quicksort algorithm.

***Sequential Recursive Merge Sort Algorithm***

*mergesort(S, low, high)*

1. *If low < high, then*
   a. *mid = (low + high)/2*
   b. *mergesort(S, low, mid-1)*
   c. *mergesort(S, mid, high)*
   d. *merge(S, low, high, mid)*

*(a)*

***Sequential Non-Recursive Merge Sort Algorithm***

*non-recursive_mergesort(S, low, high)*

1. *If low < high, then*
   a. *Sort each two elements together till the end of S*
   b. *Merge two sorted lists*
   c. *Repeat step (b) till has one sorted list*

*(b)*

***Sequential Quicksort Algorithm***

*quicksort(S, low, high)*

1. *If high <=1, then return*
   a. *Choose a pivot V*
   b. *Partition S around V to get $S_1$, $S_2$ of size $n_1$, $n_2$*
   c. *quicksort($S_1$, low, $n_1$-1)*
   d. *quicksort($S_2$, $n_1$, high)*

*(c)*

Figure 1. Sequential Merge sort & Quicksort algorithms

## 3.2 Shared Memory Parallel Merge sort

We implemented two algorithms for shared memory MIMD using OpenMP. Figure 2 shows the general algorithm for shared memory implementation. The two algorithms have the same overall structure using a parallel Merge sort, one uses non-recursive Merge sort to sort each partitioned data/thread while the other uses recursive Quicksort.

Shared-Parallel Non-Recursive Merge Sort: The data is divided among parallel threads, each partition is sorted using the non-recursive sort described in the previous section. Then using a parallel Merge sort where threads sort and merge two children sorted lists (its list and its neighbor's list) in a binary tree fashion. This step is repeated using half of available threads in each round till only one thread owns all sorted data. This algorithm works with a power of two number of threads. And after the first round the number of idle threads are doubled at each step.

Shared-Parallel Hybrid Quicksort and Merge Sort: In this version sequential recursive Quicksort is used to sort the data assigned to each parallel thread, then parallel Merge sort is used to merge the sorted lists into one list in parallel as described above. As the hybrid method showed the best performance (Figure 5), we used this combination for our distributed memory implementation next.

***Merge Sort Algorithm (Shared Memory)***

***Master Thread***

1. *Set threads number*
2. *Calculate load for threads*
3. *Create parallel region*

***All Threads***

1. *Get thread ID*
2. *Calculate start & end indexes for the thread (private)*
3. *Call sorting function (Quicksort OR Merge Sort)*
4. *Wait for all threads to finish sorting their parts*
5. *If thread has work, then: do (a-d)*
    a. *Duplicate data size*
    b. *Calculate end index of its part*
    c. *Call merge & sort function*
    d. *If master thread, then: Divide number of threads by two*
    e. *Repeat steps (a-d)*

Figure 2. Shared Memory Parallel sort algorithm

## 3.3 Distributed Memory Parallel Hybrid Quicksort and Merge Sort

For the distributed memory MIMD platform, we used the hybrid Quicksort and Merge sort similar to the shared memory version. Figure 3 shows the algorithm where the sequential recursive Quicksort is used to sort the data partitioned and sent to each of the parallel processes, then parallel Merge sort is used to merge the sorted lists into one list in parallel by communicating processes. The implementation is done using MPI.

***Merge Sort Algorithm (Distributed Memory)***

*Set Parallel environment*

***Master Node***

1. *Calculate load for each node*
2. *Send size and data to nodes*

***All Nodes***

1. *Receive the size and data*
2. *Sort data by calling Quicksort function*
3. *Decide if the node has to send data, receive, or stay idle*
4. *If the node has to send data, then:*
    a. *Calculate the receiver nod ID*
    b. *Send size and data to the receiver node*
5. *If the node has to receive data, then:*
    a. *Calculate the receiver node ID*
    b. *Receive size and data from the sender node*
    c. *Call merge & sort function*
6. *Repeat steps (3-5) till only one node has all data*

Figure 3. Distributed Memory Parallel Hybrid Quicksort and Merge Sort

## 3.4 Hybrid Memory Parallel Merge Sort Using Hybrid MSD-Radix and Quicksort in Cluster Platforms

Our hybrid algorithm designed for the cluster environment, takes advantage of existence of shared memory within the compute nodes and is implemented using MPI (among compute nodes) and OpenMP(within individual compute nodes). To reduce the overhead of communication between nodes (MPI), we implemented one step of MSD-Radix algorithm at the beginning of the algorithm. The master node divides the data into ten parts based on the most significant digit. Then sends each bucket or buckets to proper node. To achieve the better performance inside each node, the Shared-Parallel Hybrid Quicksort and Merge Sort is executed by parallel threads in a node, the sorted data is then sent to the master node to its place in the original array. Note that it is important to use MSD (and not LSD) Radix to preserve data locality and minimize the need for frequent data movement. In this MPI version the number of nodes can be varied from 1 to 10 and the number of threads should be a power of two. Figure 4 Shows the processes of hybrid memory parallel merge sort algorithm using MSD Radix and Quicksort.

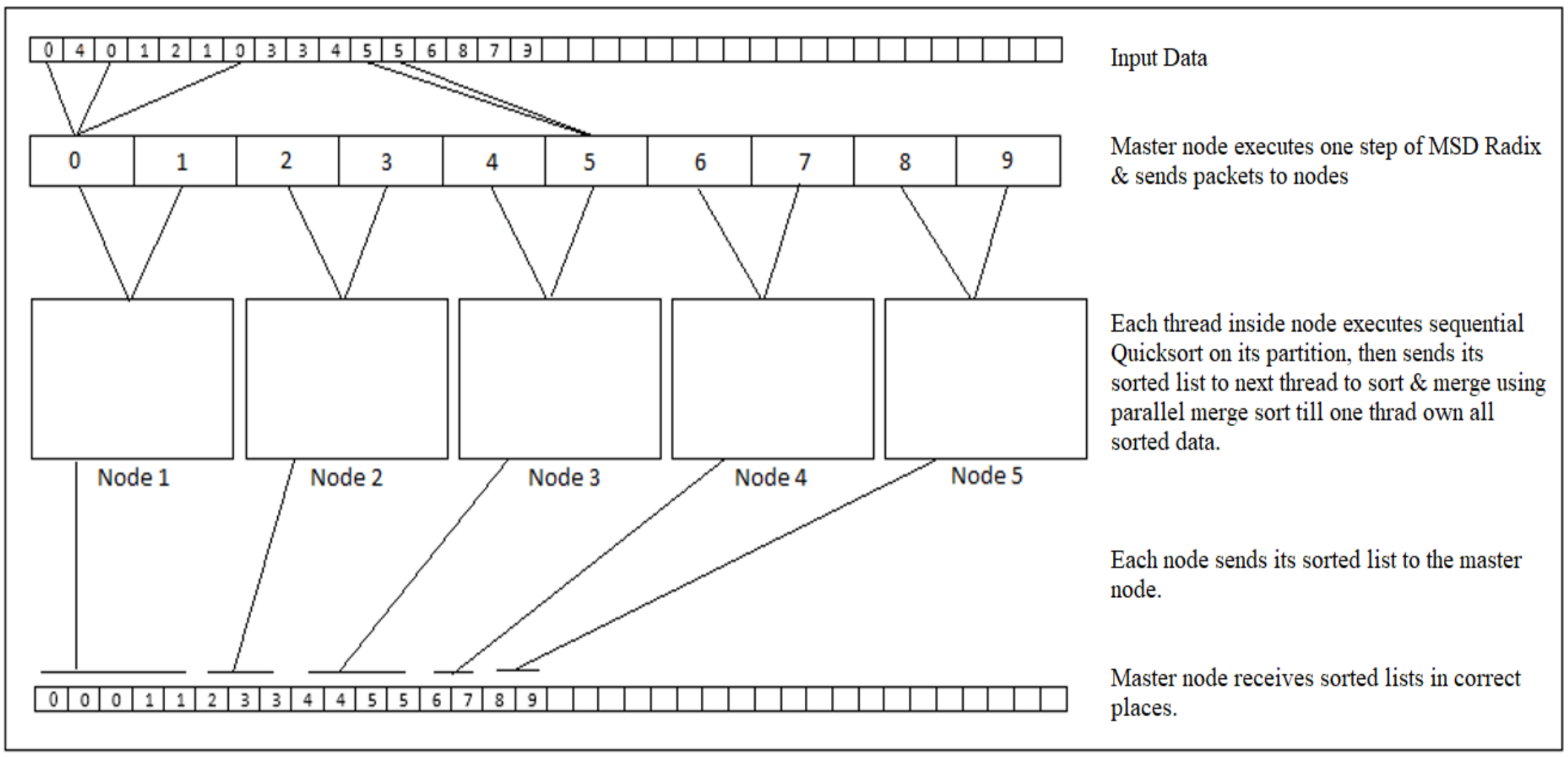

Figure 4. Hybrid Memory Parallel sort using MSD-Radix

# RESULTS AND ANALYSIS

## 4.1 Coding and Execution Environment

The code is written in C++, OpenMP, and MPI. We have tested our implementation on a 24-core shared memory compute node (2 x Intel Xeon E5-2650v4 Processors, 12 cores per Processor) of the 384-core cluster (16 compute nodes), Heracles, available in the Parallel Distributed System lab (PDS Lab http://pds.ucdenver.edu ).

## 4.2 Sequential Merge Sort and Quicksort Analysis

The performance of the sequential recursive Merge sort, non-recursive Merge sort, and recursive Quicksort is represented at Figure 5. In this case, the data used was generated randomly where each number consists of three digits. The size of data varies in the range ($1x10^6$ – $10x10^6$). This data is stored and used for all test cases for consistency of comparisons.

Figure 5 shows that non-recursive Merge sort and recursive Quicksort significantly outperformed recursive Merge sort. We observe that as the number elements to be sorted increases, Quicksort outperforms the non-recursive Merge sort.  which makes Quicksort the best choice for sorting data sequentially. In Figure 2, recursive Quicksort sort array of size ($10x10^6$) runs1.14 times faster than the non- recursive Merge sort and 1.76 times the recursive Merge sort.

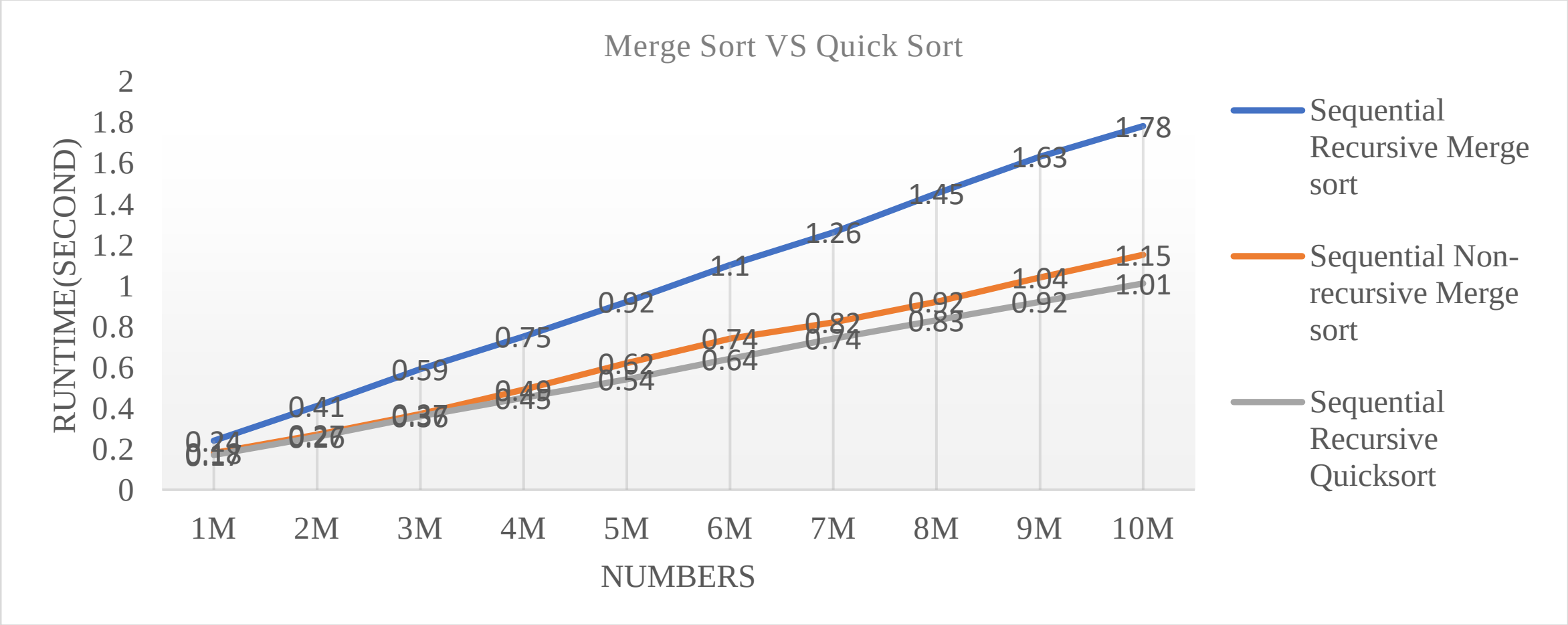

Figure 5. Performance comparison of sequential algorithms

## 4.3 Shared Memory Parallel Merge sort

We made a comparison between OpenMP versions while sorting data of size (10M) and using different number of threads. Figure 6 shows that OpenMP Shared-Parallel Hybrid Quicksort and Merge Sort outperforms OpenMP Shared-Parallel Non-Recursive Merge Sort. The speedup of OpenMP Shared-Parallel Hybrid Quicksort and Merge Sort increases significantly with the number of threads and peaks at 16 threads. We do not get more speedup at 32 threads as we are limited to 24 cores/node. There is an insignificant speedup for OpenMP Shared-Parallel Non-Recursive Merge Sort at 4 threads but then no more performance can be gained.

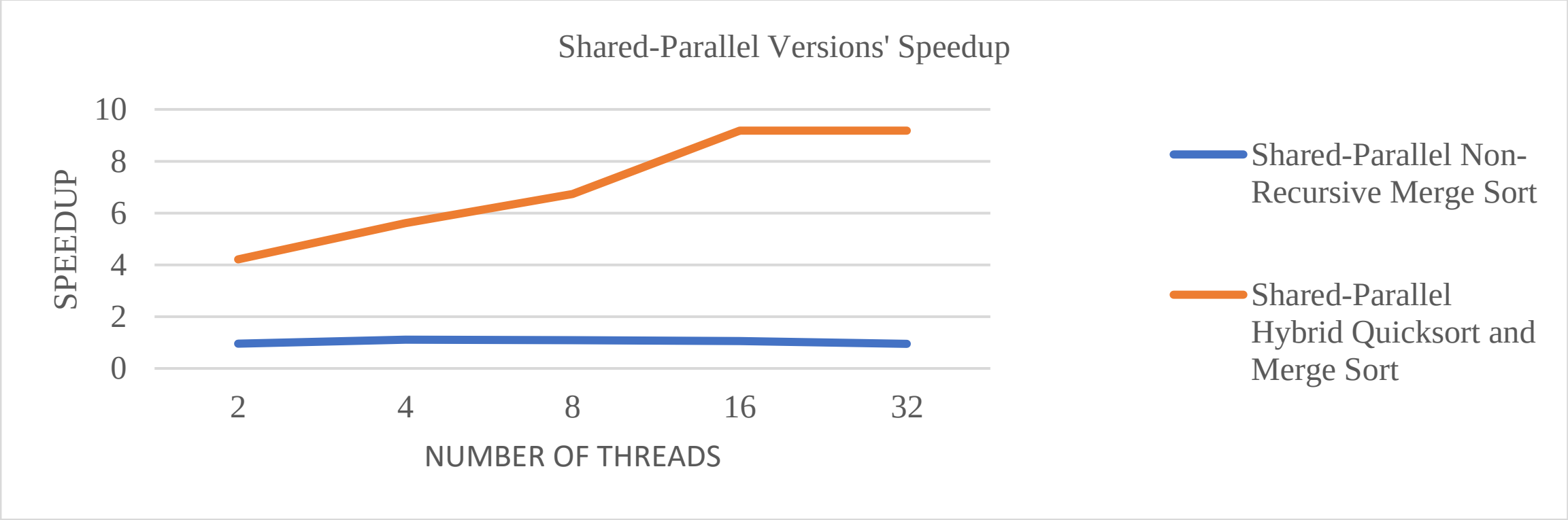

Figure 6. Performance comparison of Shared-Parallel algorithms

We also compared our OpenMP Shared-Parallel Hybrid Quicksort and Merge Sort with published fast OpenMp algorithm as baseline (*Aydin & Alaghband 2013*). The baseline algorithm implemented a hybrid parallel algorithm of non-recursive Most Significant Digit Radix MSD-Radix and Quicksort. Figure 7 shows the performance of our OpenMP Shared-Parallel Hybrid Quicksort and Merge Sort comparing with baseline algorithm using the same machine (Hydra), the same number of threads, the same data and size, and the same number of digits. Our OpenMP Shared-Parallel Hybrid Quicksort and Merge Sort shows better performance than the baseline; and continues to improve as the size of data increased. In the case of sorting 4M data using eight threads, our approach OpenMp Shared-Parallel Hybrid Quicksort and Merge Sort runs 2.55 times faster than the baseline. Merging tow sorted lists instead of repeating Quicksort for each round improved the performance of OpenMP Shared-Parallel Hybrid Quicksort and Merge Sort.

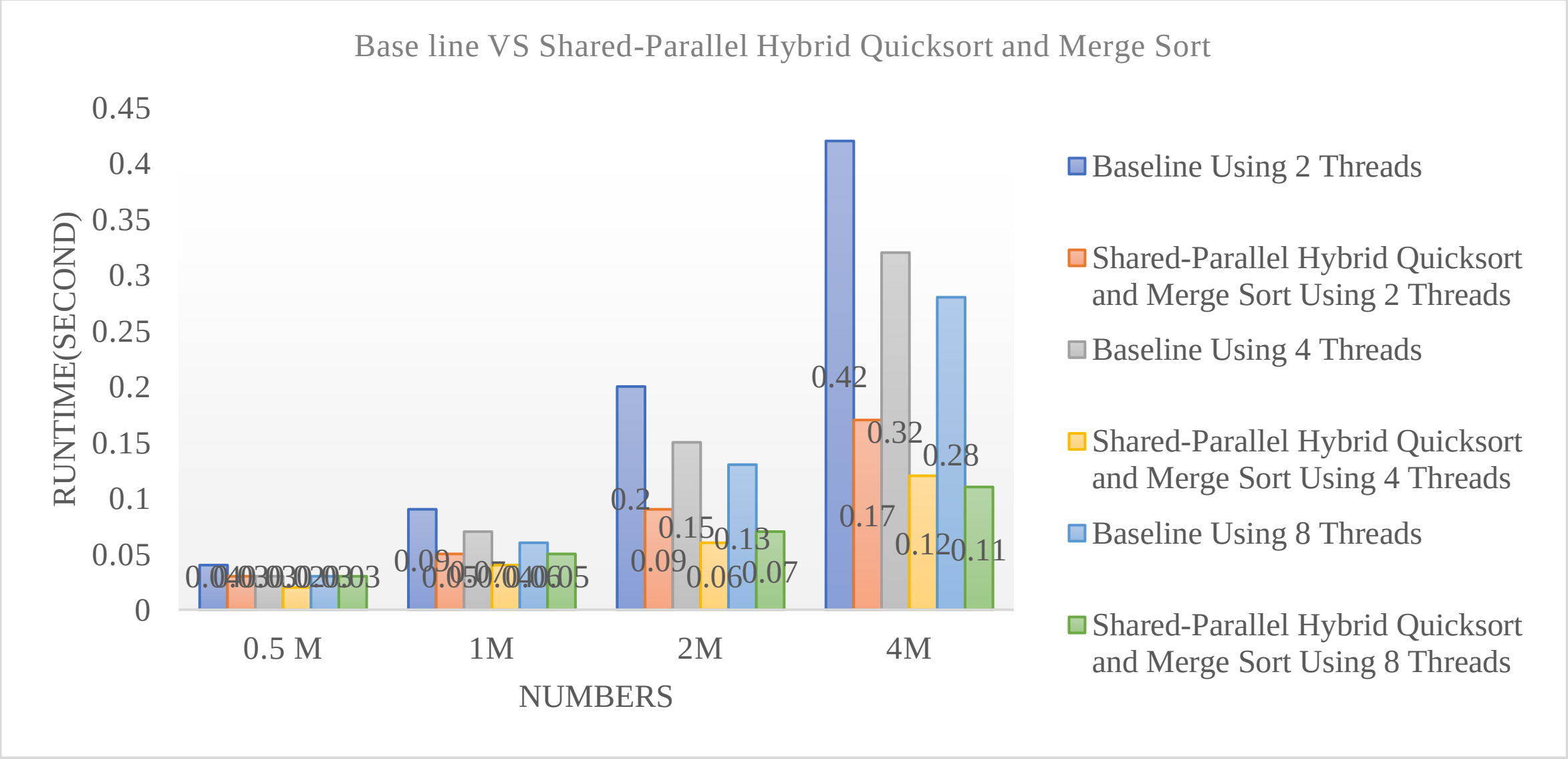

Figure 7. Performance comparison of Baseline & Shared-Parallel Hybrid Quicksort and Merge Sort OpenMP algorithms

## 4.4 Distributed Memory Parallel Hybrid Quicksort and Merge Sort

Comparison between the two OpenMP Shared-Parallel Non-Recursive Merge Sort, Shared-Parallel Hybrid Quicksort and Merge Sort and the Distributed Memory Parallel Hybrid Quicksort and Merge Sort using four threads and four nodes is shown in Figure 8. We observe that MPI algorithm outperformed OpenMP Shared-Parallel Hybrid Quicksort and Merge Sort especially when size of data is increased. In the case of sorting 10M elements, MPI algorithm runs 1.04 faster than OpenMP Shared-Parallel Hybrid Quicksort and Merge Sort. The overhead related to communication between nodes in MPI has less impact compared to data fitting inside the RAM memory in OpenMp where many long latency memory accesses must made.

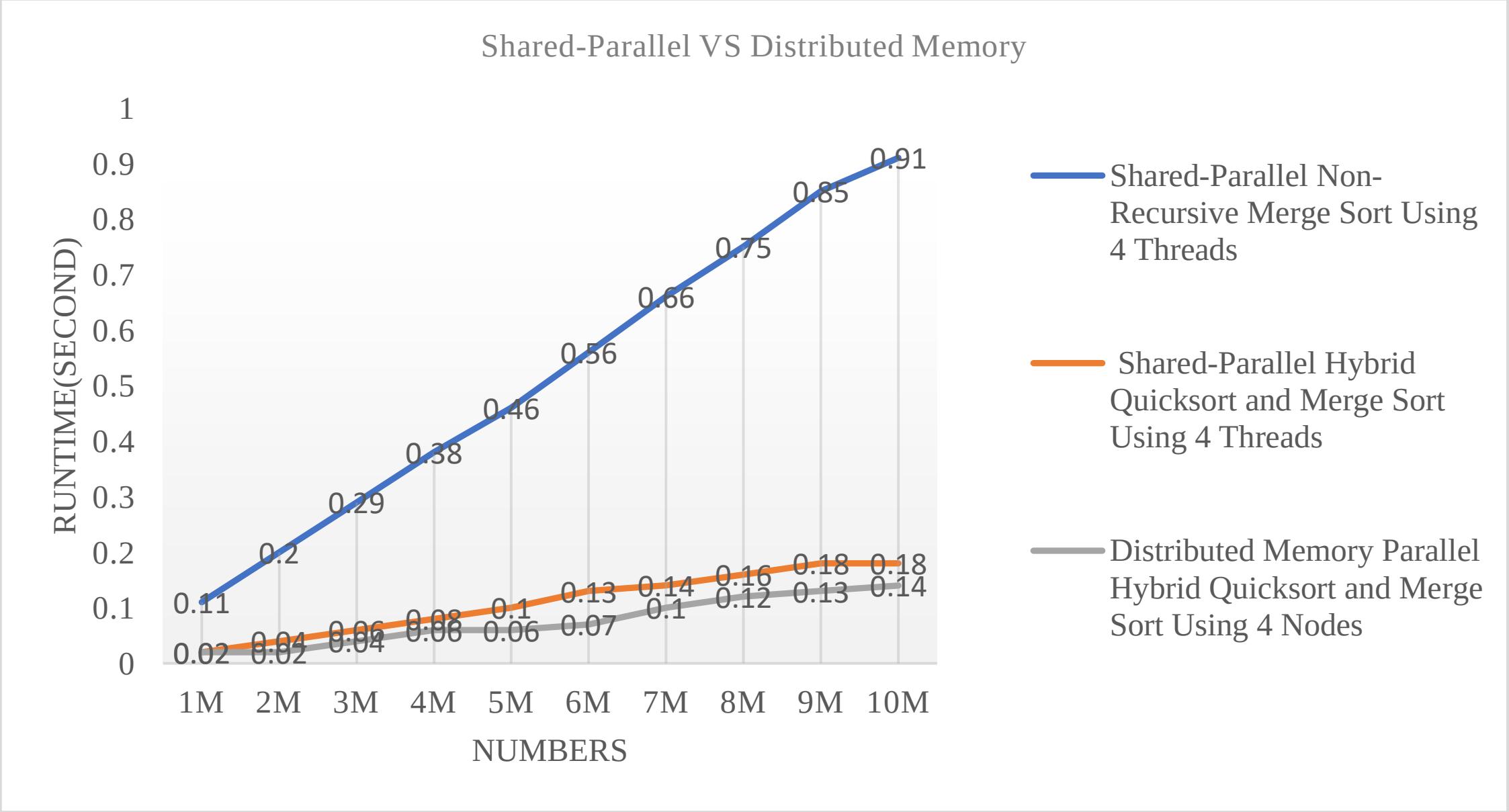

Figure 8. Performance comparison of Shared-Parallel Non-Recursive Merge sort, Shared-Parallel Hybrid Quicksort and Merge sort and the Distributed Memory Parallel Hybrid Quicksort and Merge sort

## 4.5 Hybrid Memory Parallel Merge Sort Using Hybrid MSD-Radix and Quicksort in Cluster Platforms

Figure 9 shows the speedup gains while sorting data for various array sizes by running four models; the OpenMP Shared-Parallel Non-Recursive Merge Sort, Shared-Parallel Hybrid Quicksort and Merge Sort using 4 threads, the Distributed Memory Parallel Hybrid Quicksort and Merge Sort using 4 nodes, and the Hybrid Memory Parallel Merge Sort Using Hybrid MSD-Radix and Quicksort in Cluster Platforms using two nodes and two threads. The speedup of Hybrid MPI using two nodes and two threads keeps going up with data size increases while speedup of all other parallel algorithms goes down.

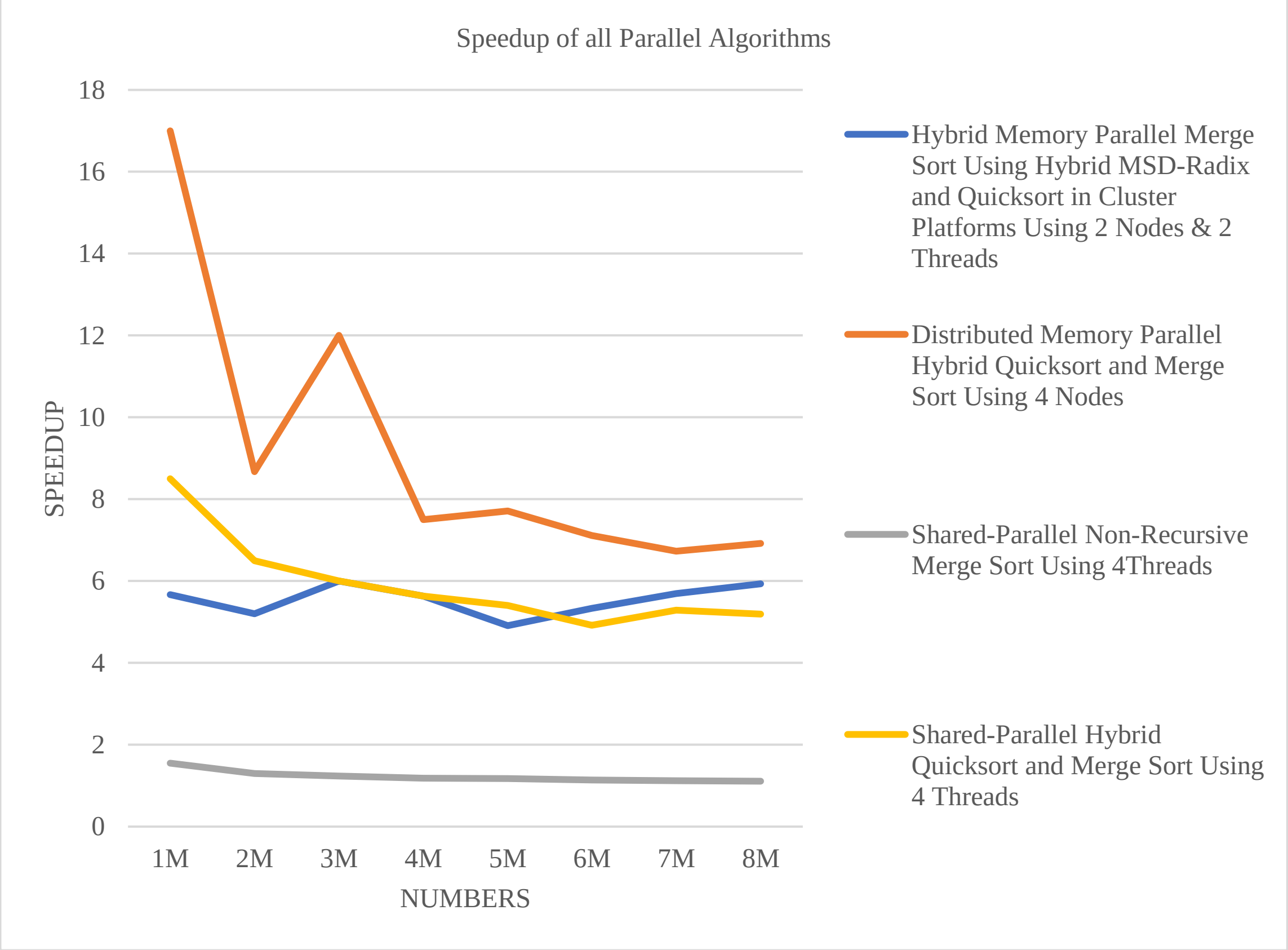

Figure 9. Performance comparison of all parallel algorithms

Figure 10 and 11 show that the speedup of Hybrid Memory Parallel Merge Sort Using Hybrid MSD-Radix and Quicksort in Cluster Platforms affected by the number of threads and nodes. Figure 10 shows the speedup of Hybrid Memory Parallel Merge Sort Using Hybrid MSD-Radix and Quicksort in Cluster Platforms in two cases, the first case used fife nodes and two threads, and the second case used fife nodes and eight threads. The second case always shows better speedup than the first case. In figure 10, we used the same number of nodes which means the same cost of communication between nodes. So, increasing the number of threads always enhances the speedup.

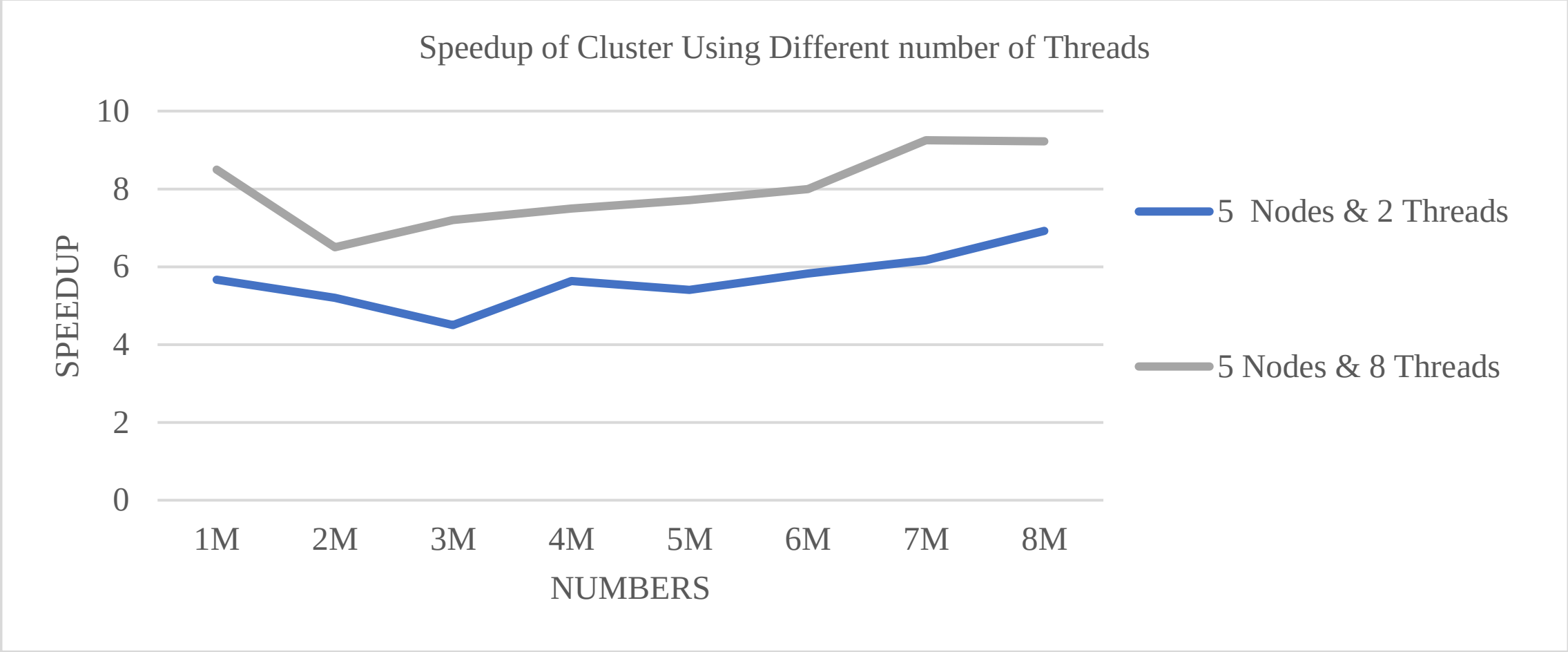

Figure 10. Speedup of cluster for different number of threads

Figure 11 shows the speedup of Hybrid Memory Parallel Merge Sort Using Hybrid MSD-Radix and Quicksort in Cluster Platforms in two cases, the first case used two nodes and two threads, and the second case used ten nodes and two threads. The first case shows better speedup than the second case when the data size less than 4M, but when the data size becomes larger than 4M the second case outperformed. In figure 11, using more nodes means more communication between nodes and more cost. But after 4M increasing the number of nodes worth because the communication overhead starts losing its effect and the workload has the significant impact now.

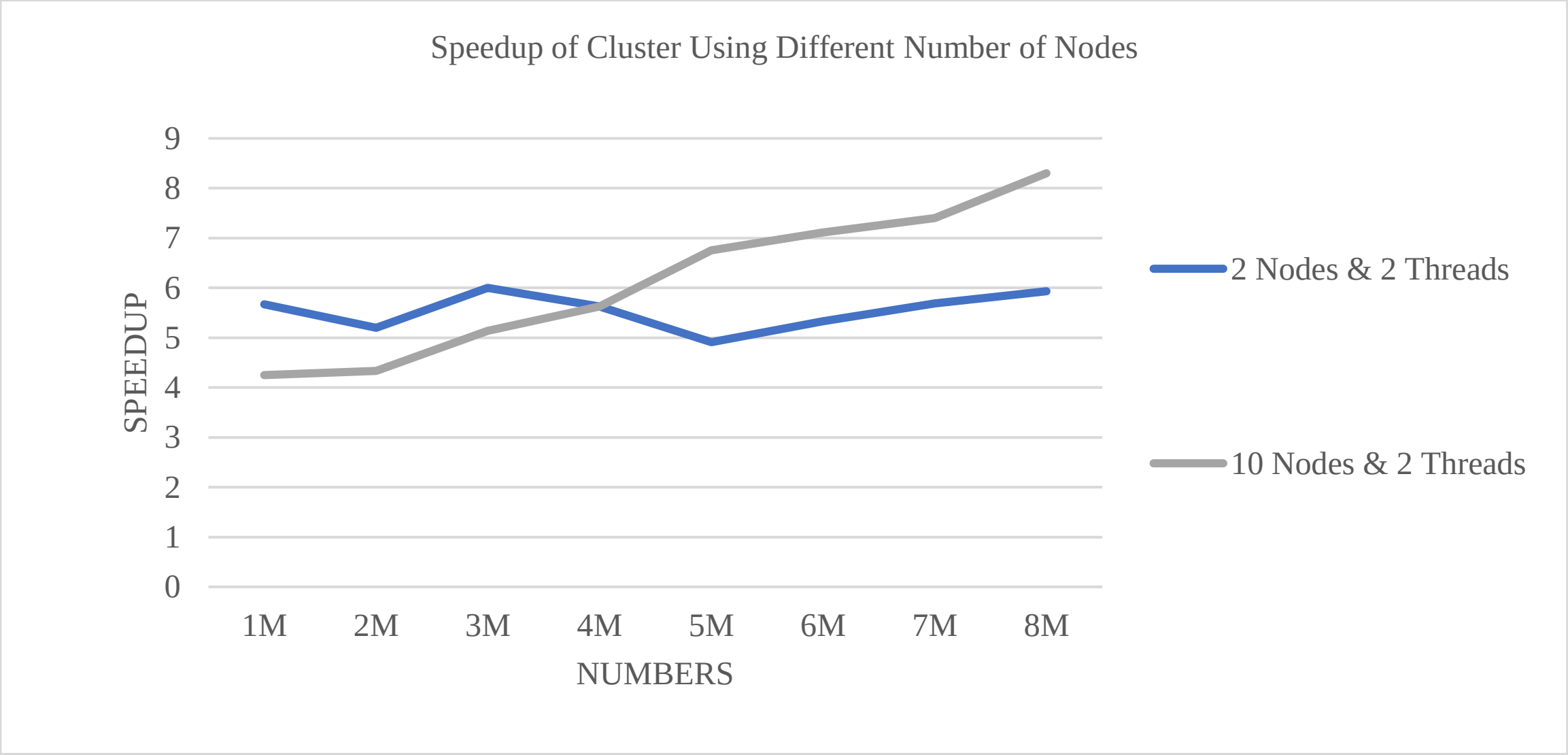

Figure 11. Speedup of cluster for different number of nodes

## Conclusion

Three parallel models for sorting are implemented on three different platforms; shared memory, distributed memory, and hybrid memory. The shared memory and distributed memory implementation used hybrid algorithms of Merge sort and Quicksort. The cluster implementation employed one step of MSD-Radix sort and applied the shared memory algorithm inside the node and applied the distributed memory algorithm between the nodes. Results presented, show that all three parallel models perform significantly better than the faster sequential Quicksort. The Shared-Parallel Hybrid Quicksort and Merge Sort performs remarkably better than the baseline algorithm. The Distributed Memory Parallel Hybrid Quicksort and Merge Sort performs better than the Shared-Parallel Hybrid Quicksort and Merge Sort. The Hybrid Memory Parallel Merge Sort Using Hybrid MSD-Radix and Quicksort in Cluster Platforms performance outperforms the Shared-Parallel Hybrid Quicksort and Merge Sort when runs on large data and its speedup keeps improving.

# REFERENCES

Al-Dabbagh, S., & Barnuti, N., 2016. Parallel Quicksort Algorithm using OpenMP. International Journal of Computer Science & Mobile Computing, Vol. 5, No. 6, pp 372-382.

Alyasseri, Z., et al, 2014. Parallelize Bubble and Merge Sort Algorithms Using Message Passing Interface (MPI). *arXiv preprint arXiv: 1411.5283*, 2014, [online] Available: https://arxiv.org/ftp/arxiv/papers/1411/1411.5283.pdf

Aydin, A., & Alaghband, G., 2013. Sequential & Parallel Hybrid Approach for Non-Recursive Most Significant Digit Radix Sort. *10th International Conference on Applied Computing*. Fort Worth, USA, pp. 51-58.

Chapman, B. et al, 2007. *Using OpenMP - Portable Shared Memory Parallel Programming*. The MIT Press, Cambridge, USA.

Durad, M., et al, 2014. Performance Analysis of Parallel Sorting Algorithms using MPI. *12th International Conference on Frontiers of Information Technology.* Islamabad, Pakistan, pp. 202-207

El-Nashar, A., 2011. Parallel Performance of MPI Sorting Algorithms on Dual-Core Processor Windows-Based Systems. *International Journal of Distributed and Parallel Systems (IJDPS),* Vol.2, No.3.

Jordan, H., Alaghband, G., 2003. *Fundamentals of Parallel Processing,* Prentice Hall Professional Technical Reference, Upper Saddle River, USA.

Hijazi, S., & Qatawneh, M., 2017. Study of Performance Evaluation of Binary Search on Merge Sorted Array Using Different Strategies. *I.J. Modern Education and Computer Science*, pp. 1-8

Kavi, k., et al, 2000. Shared Memory and Distributed Shared Memory Systems: A Survey. *Advances in computer*. Vol. 53, pp. 55-108

Kim, E., & Park, K., 2009. Improving multikey Quicksort for sorting strings with many equal elements. *Information Processing Letters*, Vol. 109, No. 9, pp. 454-459.

PDS Lab 2019. Available from < http://pds.ucdenver.edu/>

Radenski, A., 2011. Shared Memory, Message Passing, and Hybrid Merge Sorts for Standalone and Clustered SMPs. *The 2011 International Conference on parallel and distributed processing techniques and applications*. Las Vegas, USA, pp. 367-373.

Zurek, D., et al. 2013. The Comparison of Parallel Sorting Algorithms Implemented on Different Hardware Platforms. *Computer Science Journal*, Vol. 20, No. 2, pp 679-691.